# Observation of non-volatile anomalous Nernst effect in altermagnet with collinear Néel vector

Lei Han[1]†, Xizhi Fu[2]†, Wenqing He[3]†, Yuxiang Zhu[1], Jiankun Dai[1], Wenfeng Yang[1], Wenxuan Zhu[1], Hua Bai[1], Chong Chen[1], Caihua Wan[3], Xiufeng Han[3], Cheng Song[1]*, Junwei Liu[2]*, and Feng Pan[1]*

[1] Key Laboratory of Advanced Materials (MOE), School of Materials Science and Engineering, Tsinghua University, Beijing 100084, China.

[2] Department of Physics, The Hong Kong University of Science and Technology, Hong Kong 999077, China.

[3] Beijing National Laboratory for Condensed Matter Physics, Institute of Physics, Chinese Academy of Sciences, Beijing 100190, China.

†These authors contributed equally to this work.

*Corresponding author. Email: songcheng@mail.tsinghua.edu.cn; liuj@ust.hk; panf@mail.tsinghua.edu.cn.

**Anomalous Nernst effect (ANE), a widely investigated transverse thermoelectric effect that converts waste heat into electrical energy with remarkable flexibility and integration capability, has been extended to antiferromagnets with non-collinear spin texture recently. ANE in compensated magnet with collinear Néel vector will bring more opportunities to construct magnetic-field-immune and ultrafast transverse thermoelectric converters, but remains unachieved for long. It is due to the degenerated band structure of traditional collinear compensated magnet excludes non-zero Berry curvature. Here, we realize non-volatile ANE in altermagnet $Mn_5Si_3$ thin film with collinear Néel vector, whose unique alternating spin-splitting band structure plays vital role in creating non-zero Berry curvature and hotpots of anomalous Nernst conductivity near band intersections.**

**Interestingly, ANE is relatively weak in stoichiometric Mn$_5$Si$_3$, but undergoes a sixfold enhancement through strategically raising the Fermi level by additional Mn doping, indicating sensitive intrinsic influence from specific location of the Fermi level on ANE in altermagnet. Moreover, our investigation reveals a unique Néel-vector-dependent temperature-scaling relationship of anomalous Nernst conductivity in Mn$_5$Si$_3$. Our work not only fills a longstanding gap by confirming the presence of non-volatile ANE in collinear compensated magnet, but also enlightens thermoelectric physics related to exotic spin-splitting band structure in altermagnet.**

Anomalous Nernst effect (ANE), as a transverse thermoelectric conversion effect, possesses inherent advantages such as large-area-integration capability with flexible waste heat source, distinguishing from its longitudinal counterpart of the Seebeck effect (*1-4*). Over the past few decades, ANE has been widely investigated in a variety of materials, including traditional ferromagnets and ferromagnetic multilayers (*5-10*), Heusler alloys (*11-14*), 2D magnetic materials and topological materials (*1, 3, 15-22*). Recently, the observation of ANE was extended to antiferromagnet (AFM) (*23-27*). Non-collinear antiferromagnetic spin texture are required for the state-of-art ANE in AFMs, such as inverse triangular spin structure of Mn$_3$Sn (*23*) and canted spin configuration of YbMnBi$_2$ (*25*), to produce non-zero Berry curvature. However, non-collinear antiferromagnetic spin texture may lead to a reduction in intrinsic frequency [Mn$_3$Sn, chiral-spin rotation frequency of only several gigahertz (*28, 29*)] and volatile ANE under zero magnetic field (YbMnBi$_2$), which loses the ultrafast advantage of AFM (*30, 31*) and blocks practical applications with requirements of zero magnetic field. In comparison, AFM with collinear antiparallel spin configuration, i.e., collinear Néel vector, features ideal terahertz dynamics, exhibiting more potential for practical ultrafast thermoelectric converters based on non-volatile ANE. Therefore, a longstanding but unachieved goal is realizing ANE in compensated magnets with collinear Néel vector, particularly yielding thermoelectric voltage under zero magnetic field.

The core difficulty lies in that, the electronic bands in traditional AFMs with collinear Néel vector possess Kramer's degeneracy, which leads to zero total Berry curvature and subsequent no ANE. Recently, altermagnets (*32-34*), characterized by *C*-paired spin-momentum locking with broken *PT* symmetry or $T \cdot t_{1/2}$ symmetry (*35*), have raised extensive research interest (*35-44*). Numerous novel non-relativistic physical phenomena have been unveiled in these materials, including spin-splitting torque and its Onsager reciprocity effect (*45-49*), tunneling and giant magnetoresistance (*50, 51*), and piezomagnetism (*35*). We notice that, these materials enable non-relativistic lifting of Kramer's degeneracy (*52-56*) while simultaneously preserve collinear Néel vector and compensated moments (*32-34*), and thus are very promising to achieve the above goal.

The alternating spin-splitting band structure in altermagnets plus spin-orbit coupling (SOC) can generate non-trivial Berry curvature, giving rise to anomalous Hall effect (AHE) (*57-60*), but AHE does not guarantee a large ANE. It is because the former is determined by the Berry curvature for all the occupied bands (Fig. 1A and 1B, denoted by green shadow), whereas the latter originates from the Berry curvature around the Fermi level (Fig. 1A and 1C, denoted by purple shadow) (*5, 61*). As a result, investigating ANE in altermagnets can not only answer whether ANE exists in material with collinear Néel vector, but also reveal rich emergent physics related to the alternating spin-splitting band structure near the Fermi level of altermagnet.

In this work, we present the observation of non-volatile ANE in altermagnet $Mn_5Si_3$ thin film with collinear Néel vector, whose spin-splitting band structure contributes to nonzero anomalous Nernst conductivity (ANC) near band intersections when considering spin-orbit coupling. Surprisingly, ANC is initially weak for stoichiometric $Mn_5Si_3$ thin film but undergoes a drastic sixfold enhancement for small amount Mn doping film of $Mn_{5.10}Si_{2.90}$, attributed to the shift in the Fermi level. Furthermore, $Mn_5Si_3$ has a unique Néel-vector-determined temperature-scaling relationship of ANC, elucidated through a self-consistent phenological model combined with Monte-Carlo simulations.

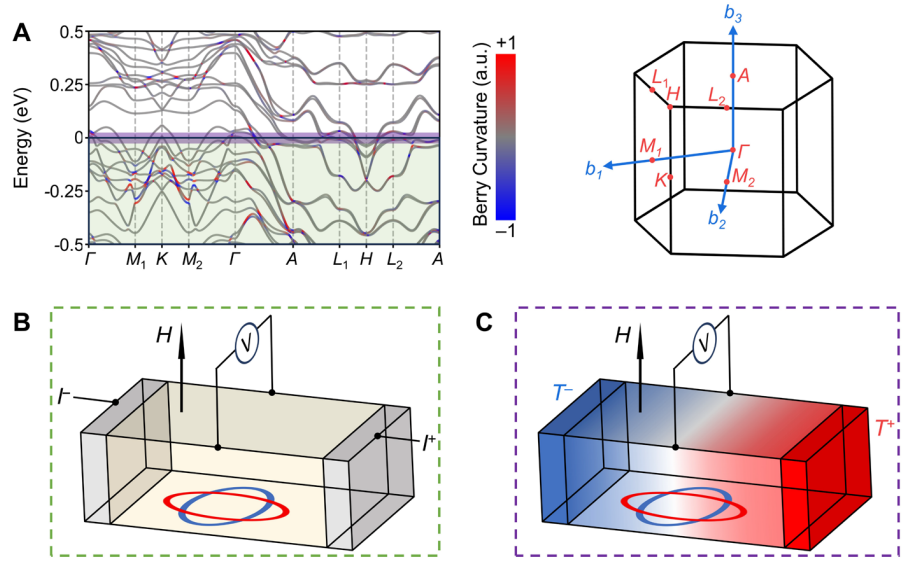

**Fig. 1. Different physical origin of AHE and ANE in altermagnet Mn$_5$Si$_3$ thin film.** (**A**) Calculated spin-splitting band structure along high symmetry points in the Brillion zone with Berry curvature as color mapping. SOC is taken into consideration. (**B, C**) experimental configurations for AHE and ANE measurements. Green shadow in (**A**) illustrates that the Berry curvature for all the occupied bands contribute to AHE as presented in (**B**). Purple shadow in (**A**) illustrates that only Berry curvature around Fermi level contribute to ANE as presented in (**C**).

**Structural features and magnetic phase transition properties of Mn$_5$Si$_3$ film**

Mn$_5$Si$_3$, an important altermagnet with the space group of P6$_3$/mcm at room temperature (*62*) has attracted wide attention due to its non-volatile AHE hysteresis (*59*), theoretically large tunneling magnetoresistance (*51*), electrical 180º Néel vector switching (*63*) as well as the cost-effective utilization of natural-abundant resources.

To investigate the thermoelectric physics generated by the spin-splitting band structure of altermagnet Mn$_5$Si$_3$, it is vital to obtain epitaxial Mn$_5$Si$_3$ thin film with high crystallinity. We prepare Mn$_5$Si$_3$ thin films by magnetron sputtering on Al$_2$O$_3$(0001) substrate (Materials and Methods), where strong Mn$_5$Si$_3$(0001) texture is confirmed by out-of-plane X-ray diffraction (XRD) with out-of-plane lattice parameter $c$ of 4.81 Å (fig. S1). High-angle annular dark field scanning transmission electron microscope (HAADF-STEM) image presented in Fig. 2A shows an epitaxial growth mode of Al$_2$O$_3$(0001)[01$\bar{1}$0]//Mn$_5$Si$_3$(0001)[11$\bar{2}$0]. In essence, the Mn$_5$Si$_3$ crystal is 30º-rotated epitaxially grown on the Al$_2$O$_3$ substrate. This growth mode can be depicted by the schematic crystal structure of the Al$_2$O$_3$(0001)/Mn$_5$Si$_3$(0001) heterojunction (Fig. 2B), highly consistent with the region marked by green box in the HAADF-STEM image of Fig. 2A. Further scrutiny through a magnified HAADF-STEM image (Fig. 2C) consolidates the high crystalline quality of the Mn$_5$Si$_3$ film, showcasing a clear atomic resolution of Mn atoms. Mn$_a$ and Mn$_b$ locate at two inequivalent Wyckoff positions (51). Due to the 30º-rotated growth mode, 3 Mn$_5$Si$_3$ unit cells matches 5 Al$_2$O$_3$ unit cell along the Mn$_5$Si$_3$[11$\bar{2}$0] (or Al$_2$O$_3$[01$\bar{1}$0]) direction (Fig. 2B), where corresponding epitaxial match relationship between interfacial Mn$_a$ and interfacial Al atoms is depicted in Fig. 2D. High crystalline quality and 30º-rotated epitaxial growth mode are also validated by $\varphi$-scan XRD and HAADF-STEM image of another zone axis (fig. S2).

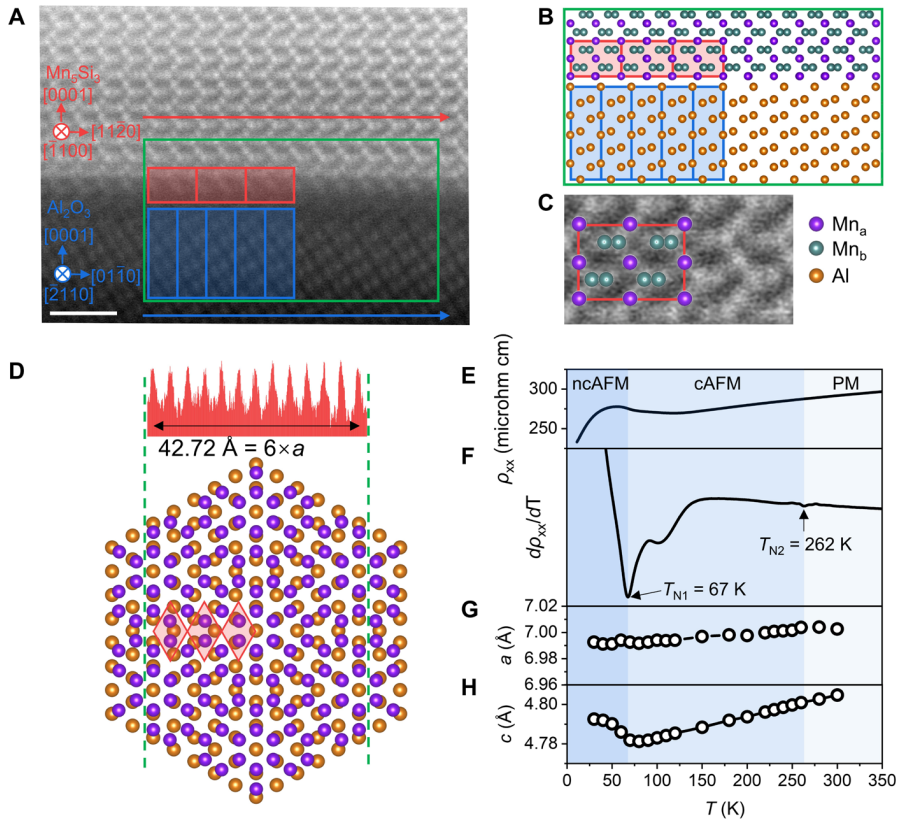

**Fig. 2. Structural characterization and magnetic phase transition properties of Mn$_5$Si$_3$ thin film.** (**A**) HAADF-STEM image of Al$_2$O$_3$(0001)/Mn$_5$Si$_3$(0001) heterojunction under zone axis of Al$_2$O$_3$<$\bar{2}$110>. The scale bar denotes 1 nm. (**B**) Crystal structure of Al$_2$O$_3$(0001)/Mn$_5$Si$_3$(0001) heterojunction for the region denoted by green box in (**A**). Red and blue boxes mark the unis cells for Mn$_5$Si$_3$ and Al$_2$O$_3$, respectively, which are consistently denoted in the same color in Fig. 2a and Fig. 2d. (**C**) A magnified HAADF-STEM image of Mn$_5$Si$_3$, where the unit cell as well as Mn$_a$ and Mn$_b$ atoms are denoted. (**D**) Schematic epitaxial match relationship between interfacial Mn$_a$ and interfacial Al atoms. (**E**) Temperature-dependent longitudinal resistivity curve and (**F**) its temperature derivative of the Mn$_5$Si$_3$ thin film. (**G**) Temperature-dependent lattice parameter $a$ and (**H**) lattice parameter $c$ of the Mn$_5$Si$_3$ thin film.

The epitaxial growth of Mn$_5$Si$_3$ thin film brings about epitaxial strain. The in-plane lattice parameter $a$ of Mn$_5$Si$_3$ is measured to be 7.12 Å, by averaging the total interplanar spacing of 6

Mn$_5$Si$_3$ unit cells along the red arrow denoted in Fig. 2A. Compared with $a$ of 6.91 Å for bulk Mn$_5$Si$_3$ (*62*), a larger $a$ for our thin film Mn$_5$Si$_3$ indicates an in-plane elongation of the unit cell. Such an in-plane tensile strain of Mn$_5$Si$_3$ thin film can also be cross-checked by macroscopic in-plane XRD measurements (fig. S1), consistent with the microscopic characterization of HAADF-STEM.

Epitaxial strain strongly influences crystal lattice of Mn$_5$Si$_3$ thin film, which tunes its altermagnetic local spin environment and corresponding emergent physical properties (*39, 44*). One typical example is that the magnetic phase transition of Mn$_5$Si$_3$ thin film is very different compared to bulk Mn$_5$Si$_3$. For bulk Mn$_5$Si$_3$, there are two magnetic phase transition points when decreasing temperature, corresponding to paramagnet (PM) to collinear antiferromagnet (cAFM) transition ($T_{N2}$) and cAFM to non-collinear antiferromagnet (ncAFM) transition ($T_{N1}$) (*62-65*). For our thin film, we measured temperature-dependent longitudinal resistivity curve ($\rho_{xx}$-$T$) to investigate the magnetic phase transition property, where the derivative of $\rho_{xx}$-$T$ also indicates two magnetic phase transition points of $T_{N1}$ and $T_{N2}$ (Fig. 2E and 2F). Interestingly, compared with the bulk Mn$_5$Si$_3$, $T_{N2}$ in thin film Mn$_5$Si$_3$ drastically increases due to epitaxial strain, which is consistent with former measurements on thin film Mn$_5$Si$_3$ (*59, 63*).

Another difference between bulk and thin film Mn$_5$Si$_3$ lies in distinct structural transition properties. For bulk Mn$_5$Si$_3$, there are two structural transitions accompanied with magnetic phase transitions at $T_{N1}$ and $T_{N2}$, which can be inferred by abrupt changes in temperature-dependent in-plane lattice parameter $a$ and out-of-plane lattice parameter $c$ (*62, 64, 65*). In comparison, when cooling down Mn$_5$Si$_3$ thin film, $a$ decreases slightly and monotonically without abrupt changes (Fig. 2G, fig. S1), indicating the absence of structural transitions. Interestingly, Fig. 2H shows that $c$ experiences monotonic decrease when cooling down until $T_{N1}$, and occurs non-monotonic change at $T_{N1}$ (fig S1). These temperature-dependent in-plane and out-of-plane XRD results demonstrate that the unit cell of Mn$_5$Si$_3$ thin film can still freely compress and expand along $c$- axis, but cannot deform along in-plane axis due to epitaxial strain. This constraint forces the hexagonal crystal structure at room temperature to preserve until the ncAFM temperature range, bringing about prerequisites for the generation of AHE (*59, 63*) in the cAFM phase, namely the altermagnet phase of Mn$_5$Si$_3$. Our epitaxial growth of high quality Mn$_5$Si$_3$ thin films, coupled with comprehensive characterizations on magnetic phase transitions and structural transitions, establishes a foundational platform for investigating intrinsic thermoelectric transport behaviors.

**Anomalous Nernst effect in altermagnet Mn$_5$Si$_3$ thin film with collinear Néel vector**

Resulting from the alternating spin-splitting band structure and SOC, AHE has been observed in the cAFM phase of Mn$_5$Si$_3$(0001) thin film (*59, 63*). In this work, AHE also emerges with non-volatile hysteresis under out-of-plane magnetic field $H$ and in-plane current $I$, where the anomalous Hall resistivity $\rho_{yx}$ can be calculated by $\rho_{yx} = \frac{U_{yx}}{I}d = \frac{U^+ - U^-}{I}d$ (Fig. 3A). $U_{yx}$ and $d$ is the Hall resistance and the thickness of Mn$_5$Si$_3$ thin film. The as-measured non-volatile AHE brings about possibilities for non-zero ANE at zero magnetic field.

Before carrying out ANE measurements, a brief exploration and summary of the AHE's origin in the cAFM Mn$_5$Si$_3$ thin film are presented. Magnetization measurements (Fig. 3B) reveal that the raw magnetization $M$ can be decomposed into two parts, each stemming from different sources with different coercive fields. One part, denoted as $m$, has the same coercive field as that of the AHE hysteresis, attributed to weak spin canting of collinear antiparallel sublattice moments induced by relativistic Dzyaloshinskii-Moriya interaction (DMI) (*63, 66*). Notably, this DMI-induced $m$ is as small as 0.0013 $\mu_B$ per magnetic Mn atom, indicating a negligible spin canting angle of less than 0.02°, estimated from the Mn atomic moment of 2.4 $\mu_B$. The remaining part of magnetization $M-m$, with nearly zero coercive field, arises from unavoidable sputtering-induced magnetic defects, which does not contribute to anomalous Hall resistivity $\rho_{yx}$ (Fig. 3A). It is important to mention that $\rho_{yx}$ is independent of $m$, but is determined by the direction of the Néel vector *n*. Due to the fixed chirality between *m* and *n* by DMI, when the magnetic field switches *m*, *n* is also simultaneously switched to bring about the $\rho_{yx}$ hysteresis (*63*).

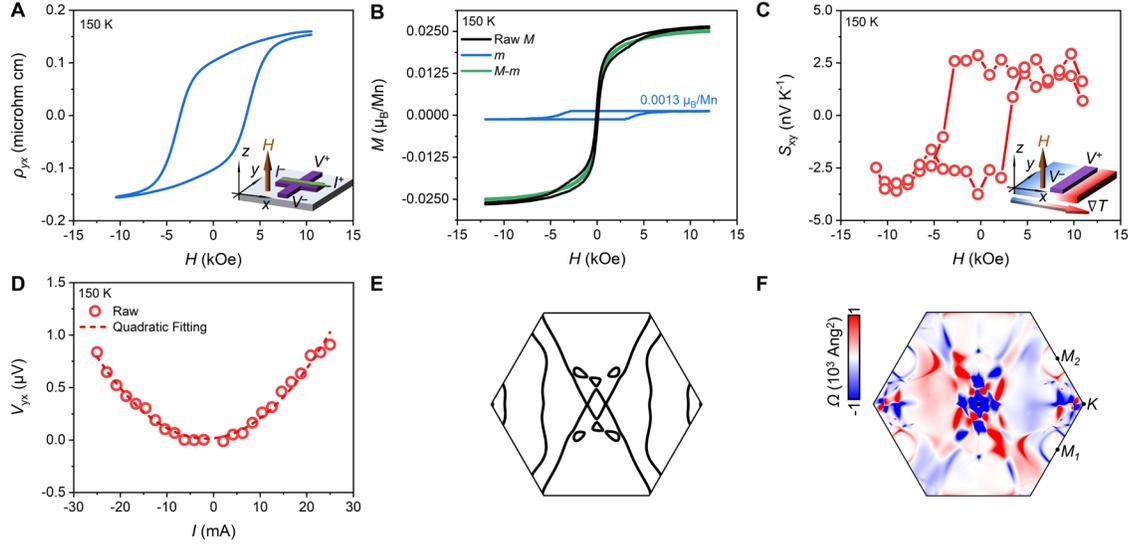

**Fig. 3. ANE generated from the spin-splitting band structure of $Mn_5Si_3$ thin film.** (**A**) Hysteresis of anomalous Hall resistivity $\rho_{yx}$ when sweeping out-of-plane magnetic field *H*. Inset illustrates the measurement configuration. (**B**) Hysteresis of magnetization *M*, which can be decomposed into DMI-induced magnetization *m* and defect induced magnetization *M-m*. (**C**) Hysteresis of anomalous Nernst coefficient $S_{xy}$ when sweeping out-of-plane magnetic field *H*. Inset illustrates the measurement configuration. (**D**) Relationship between heating current magnitude and ANE-induced transverse voltage $V_{yx}$. (**E**) Calculated spin-splitting band of $Mn_5Si_3$ at the Fermi level. With SOC, spin-up and spin-down cannot be well defined, but the spin-splitting phenomena remains. (**F**) Hotpots for ANC at *z* = 0 slice in the first Brillouin zone calculated at 150 K.

We then measure the ANE of altermagnet $Mn_5Si_3$(0001) thin film under out-of-plane magnetic field *H* and in-plane temperature gradient $\nabla T$ (inset of Fig. 3C), using analogous measurement configuration as AHE (inset of Fig. 3a) by replacing electrical current withs thermal current. By introducing electrical current *I* into Pt bars adjacent to the $Mn_5Si_3$ thin film as on-chip heaters, the in-plane temperature gradient $\nabla T$ is established. The magnitude of $\nabla T$ is quantified through monitoring the resistance of several adjacent Pt bars as thermocouples (fig. S3) (*13, 14*). The ANE-generated transverse voltage $V_{yx}$ is collected, which can be transformed into anomalous Nernst coefficient $S_{xy}$ using $S_{xy} = \frac{V_{yx}}{L(-\nabla T)}$, where $L$ is the length of the Mn$_5$Si$_3$ bar. Clear non-zero $S_{xy}$ with non-volatile hysteresis is observed (Fig. 3C), where the coercive field of $S_{xy}$ is consistent with that of $\rho_{yx}$ (Fig. 3A). The transverse voltage $V_{yx}$ demonstrates a quadratic increase when the magnitude of the heating current $I$ in the adjacent Pt bar rises, indicating that $V_{yx}$ is indeed the ANE voltage instead of AHE due to leakage current (Fig. 3D). Moreover, when the polarity of $I$ reverses, the polarity of $V_{yx}$ remains the same, which is also consistent with the physical process of ANE rather than AHE.

To unravel the physical mechanism behind the generation of non-zero ANE, we conduct first-principles calculations (Materials and Methods). When considering SOC, the non-trivial spin-splitting band structure of Mn$_5$Si$_3$ remains, exemplified by the band structure at the Fermi surface (Fig. 3E). Notably, this spin-splitting band structure results in numerous band intersections. Consequently, these band intersections serve as hotpots of ANC $\alpha_{xy}$ in the first Brillouin zone (Fig. 3F), establishing the intrinsic origin of the measured non-vanishing $S_{xy}$ upon integrating these hotpots.

**Band structure and order parameter dependence of Anomalous Nernst effect in altermagnet**

Although ANE is clearly revealed in altermagnet Mn$_5$Si$_3$ thin films, the magnitude of $S_{xy}$ is relatively small. Compared with typical antiferromagnets exhibiting non-vanishing ANE (*23, 25, 26*), $S_{xy}$ of stoichiometric Mn$_5$Si$_3$ is 2~3 orders of magnitude smaller. Surprisingly, this contrasts with the magnitude of $\rho_{yx}$, which is comparable or only slightly smaller than that of other antiferromagnets. This observation is quite unexpected, especially for our Mn$_5$Si$_3$ thin films with high crystalline quality as confirmed in Fig. 2. This phenomenon reminds us of the distinct physical origins of AHE and ANE (Fig. 1), which emphasizes that a large AHE does not guarantee a large ANE. The strong discrepancy between AHE and ANE in $Mn_5Si_3$ thin film exactly indicates intrinsic contribution from the altermagnetic spin-splitting band structure, because the location of Fermi level may accidentally lead to very small ANE for stoichiometric $Mn_5Si_3$.

As expected, the calculated $\alpha_{xy}$ at the Fermi level $E_F$ (denoted by dashed black line) is very small for stoichiometric $Mn_5Si_3$ (Fig. 4A), where the temperature for calculation has been set to be the same as experimental measurement for consistency. Besides, Fig. 4A shows that only if switching Néel vector by 180° from $n_+$ to $n_-$ can produce opposite ANC, whereas simply switching the DMI-induced net moment from $m_+$ to $m_-$ cannot. This finding supports that the ANE originates in the Néel-vector-dependent Berry curvature, rather than the net moment. More importantly, Fig. 4A reveals that $\alpha_{xy}$ will increase rapidly with an elevated $E_F$, inspiring us to experimentally introduce extra Mn into $Mn_5Si_3$ as electron doping to increase $E_F$ for achieving a higher $S_{xy}$.

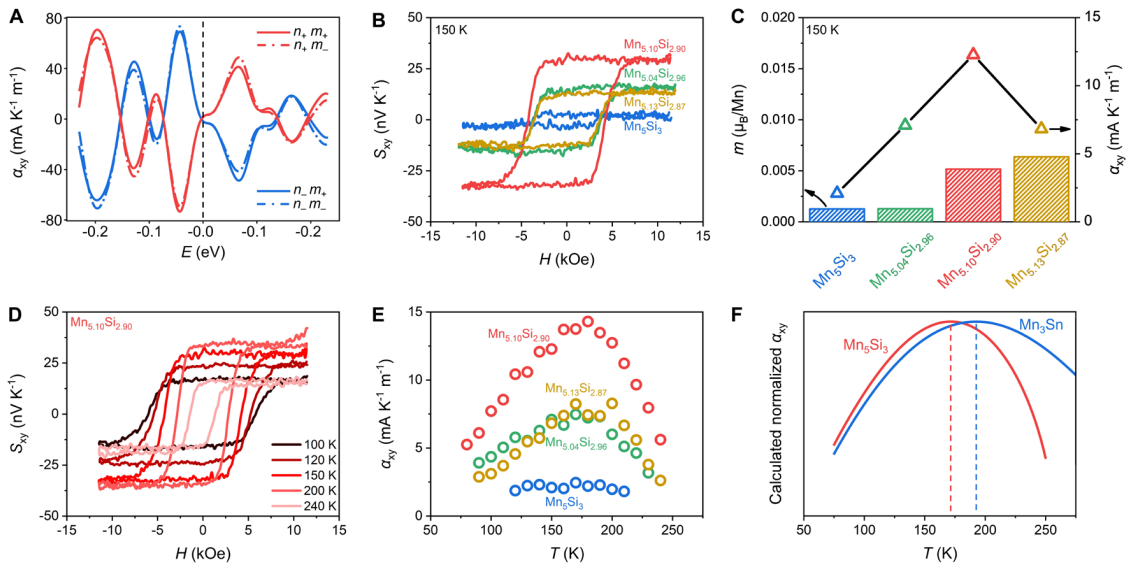

**Fig. 4. Band structure and order parameter dependence of ANE in $Mn_5Si_3$ thin film.** (**A**) Calculated relationship between ANC $\alpha_{xy}$ and energy $E$ at 150 K. (**B**) Different anomalous Nernst coefficient $S_{xy}$ for $Mn_5Si_3$ with different doping doses at 150 K. (**C**) Dependence of $m$ and $\alpha_{xy}$ with doping doses at 150 K. (**D**) Hysteresis of anomalous Nernst coefficient $S_{xy}$

when sweeping out-of-plane magnetic field $H$ under different temperatures for $Mn_{5.10}Si_{2.90}$. (**E**) Temperature dependence of $\alpha_{xy}$ for $Mn_{5+x}Si_{3-x}$ with different doping doses. (**F**) Calculated temperature dependence of normalized $\alpha_{xy}$ for altermagnet $Mn_5Si_3$ and non-collinear antiferromagnet $Mn_3Sn$.

Three types of doping samples were prepared, denoted as $Mn_{5.04}Si_{2.96}$, $Mn_{5.10}Si_{2.90}$, and $Mn_{5.13}Si_{2.87}$, respectively. The magnetic phase transition properties remain in these doping samples (fig. S4). As shown in Fig. 4B, with increasing Mn component, $S_{xy}$ initially increases and then decreases. This behavior indicates the presence of several competing factors influencing $S_{xy}$, with opposite dependence on the Mn component. To delve deeper, we measure the magnetization $m$ for each sample and compared it with $\alpha_{xy}$ (Fig. 4C), calculated using $\alpha_{xy} = S_{xy}\sigma_{xx} + \sigma_{xy}S_{xx}$. Here, $\sigma_{xx}$, $\sigma_{xy}$, and $S_{xx}$ denote longitudinal conductivity, anomalous Hall conductivity, and Seebeck coefficient, respectively (fig. S5). It is clear that there is no simple positive correlation between $\alpha_{xy}$ and $m$, providing further evidence that ANE does not arise from $m$.

These doping dependence of ANE in $Mn_5Si_3$ can be understood by considering both intrinsic and extrinsic effect. Slight Mn doping in $Mn_{5+x}Si_{3-x}$ increases $E_F$ of 0.02, 0.05, and 0.07 eV for $x =$ 0.04, 0.10, and 0.13 (Materials and Methods), dominating the intrinsic enhancement of ANE for $Mn_{5.04}Si_{2.96}$ and $Mn_{5.10}Si_{2.90}$ (Fig. 4A). Meanwhile, additional Mn doping leads to lower crystal quality, dominating the extrinsic reduction of ANE for $Mn_{5.13}Si_{2.87}$. It is worth noting that $\alpha_{xy}$ of $Mn_{5.10}Si_{4.90}$ increased approximately sixfold compared to that of $Mn_5Si_3$, unraveling the high sensitivity of ANE to the location of $E_F$ in altermagnet with spin-splitting band structure.

Furthermore, we carried out temperature-dependent ANE measurements, and the typical results for $Mn_{5.10}Si_{2.90}$ is shown Fig. 4D. As temperature decreases, $\alpha_{xy}$ first increases and then decreases (Fig. 4E), consistent with former results of ANE in antiferromagnets (*23, 25, 26*). This behavior can be qualitatively described by a self-consistent phenological toy model (text S1, fig.

S6). Nevertheless, it is important to note that the temperature derivative for temperature-dependent $α_{xy}$ varies among Mn$_3$Sn and Mn$_5$Si$_3$. Specifically speaking, as the temperature decreases, $α_{xy}$ increases rapidly and then decreases slowly (denoted as Type I $α_{xy}$-$T$) for Mn$_5$Si$_3$, while it increases slowly and then decreases rapidly (denoted as Type II $α_{xy}$-$T$) for Mn$_3$Sn (*23*).

To gain a deeper understanding of these observed differences among temperature-dependent ANE, Monte-Carlo simulations were employed, where the influence of thermal fluctuations on the order parameter, i.e., the Néel vector, is found to be important. When combined with our phenological model and considering corresponding Néel temperature of Mn$_5$Si$_3$ and Mn$_3$Sn (text S2, fig. S7), we are able to capture the detailed features of Type I $α_{xy}$-$T$ and Type II $α_{xy}$-$T$ (Fig. 4F). As a result, the relatively low Néel temperature of Mn$_5$Si$_3$ (~260 K) results in Type I $α_{xy}$-$T$, while the relatively high Néel temperature of Mn$_3$Sn (~430 K) (*23*) results in Type II $α_{xy}$-$T$. This rule also applies to YbMnBi$_2$ (*25*), which has a relatively low Néel temperature of 290 K and exhibits similar behaviors of Type I $α_{xy}$-$T$ as Mn$_5$Si$_3$. Therefore, thermal fluctuations of Néel vector significantly influences the estimation of $α_{xy}$-$T$ relationship for altermagnets.

**Discussion**

We demonstrate the feasibility of achieving ANE with non-volatile hysteresis in material with collinear Néel vector, by making use of the alternating spin-splitting band structure of altermagnet Mn$_5$Si$_3$. The spin-splitting band structure induces ANE by generating non-zero Berry curvature and subsequent ANC hotpots near band intersections. Although relatively weak in stoichiometric Mn$_5$Si$_3$ due to accidentally inappropriate location of $E_F$, ANC experiences a remarkable sixfold enhancement through Mn doping. This enhancement, linked to the increase in the Fermi level due to Mn doping, underscores the intrinsic origin of ANE in altermagnet Mn$_5$Si$_3$. Moreover, we observe a unique temperature-dependent ANC in Mn$_5$Si$_3$ arisen from thermal fluctuations of the Néel vector, which is confirmed by combing a phenological model with Monte-Carlo simulations. The ANE may also be generalized to other altermagnets with spin-splitting band structure, including RuO$_2$, MnTe, and CrSb. Our work not only expands the pool of material candidates capable of exhibiting the ANE, but also deepens the understanding of anomalous thermal transport behaviors associated with alternating spin-splitting band structure near the Fermi level.

## Materials and Methods

### Sample preparation and device fabrication

80 nm Mn$_5$Si$_3$(0001) film were grown on Al$_2$O$_3$(0001) substrate by co-sputtering Mn and Si at 600 °C with a rate of 0.4 Å s$^{-1}$ under a base pressure below 5×10$^{-8}$ Torr. For ANE measurements, Mn$_5$Si$_3$ thin films were patterned to bars of 2 mm×20 μm using optical lithography combined with Ar ion milling. Then Pt bars as on-chip thermal sources were evaporated and lift-off adjacent to the Mn$_5$Si$_3$ bars, patterned by optical lithography. An in-plane thermal temperature gradient can be established when applying current in the Pt bars, and the magnitude of temperature gradient was calibrated based on principle of thermocouple temperature measurement (fig. S3). All samples were kept in a glove box with O$_2$ and H$_2$O < 0.01 parts per million to prevent degradation or oxidation.

### Characterizations

Cross-sectional high-angle annular dark-field scanning transmission electron microscopy was conducted on an FEI Titan 80-300 electron microscopy equipped with a monochromator unit, a probe spherical aberration corrector, a post-column energy filter system (Gatan Tridiem 865 ER), and a Gatan 2k slow scan CCD system, operating at 300 kV, combining an energy resolution of ~0.6 eV and a dispersion of 0.2 eV per channel with a spatial resolution of ~0.08 nm. $\rho_{xx}$, $\rho_{yx}$, and $S_{xx}$ of Mn$_5$Si$_3$ films with different doping doses were measured using commercial Physical Property Measurement System (PPMS, Quantum Design). Ordinary Hall resistivity that is linear to magnetic field was subtracted from $\rho_{yx}$. In-situ in-plane and out-of-plane XRD at low temperatures were measured at BL02U2 Beamline from Shanghai Synchrotron Radiation Facility.

Magnetic hysteresis curves were collected using a commercial superconducting quantum interference device (SQUID, Quantum Design), where the diamagnetic contribution of $Al_2O_3(0001)$ substrate was subtracted.

**Tight binding model calculations**

The initial Hamiltonian $H_{FPLO}$ considering atomic magnetic moment and spin-orbit coupling was calculated on FPLO software (67). Then the spinless Hamiltonian $H_0$ was obtained as $H_0 = \left(H_{FPLO}^{\uparrow\uparrow} + H_{FPLO}^{\downarrow\downarrow}\right)/2$, where $H_{FPLO}^{\uparrow\uparrow}$ ($H_{FPLO}^{\downarrow\downarrow}$) is the spin-up (spin-down) part of $H_{FPLO}$. Meanwhile, the Hamiltonian contributed by the atomic magnetic moment is approximately a localized uniform magnetic field $H_Z = E_Z m_i \cdot \sigma$, where $E_Z$ is the Zeeman energy, $m_i$ is the atomic magnetic moment of $i$-th atom in the unit cell, and $\sigma$ is the Pauli matrix. As usual, the spin-orbit coupling Hamiltonian was set as $H_{SOC} = E_{SOC} L \cdot \sigma$, where $E_{SOC}$ is the spin-orbit coupling energy, $L$ the orbital angular momentum operator. Then, we can construct our final Hamiltonian $H = H_0 + H_Z + H_{SOC} = H_0 + E_Z m_i \cdot \sigma + E_{SOC} L \cdot \sigma$. Here $E_Z$ was fitted to be around −1 eV from $H_{FPLO}$, and $E_{SOC}$ was designed as 0.06 eV. Therefore, anomalous Nernst conductivity $\alpha_{xy}$ can be computed by integrating Berry curvature $\Omega_n^z(k)$ with entropy density over first Brillouin zone (5, 61), i.e.,

$$\alpha_{xy}(\mu, T) = k_B \frac{e}{\hbar} \int \frac{d^3k}{(2\pi)^3} \sum_n \Omega_n^z(k) s(\varepsilon_n(k)) \quad (1)$$

$$s(\varepsilon) = \frac{\varepsilon-\mu}{k_B T} f(\varepsilon) + \log\left(1 + e^{-\frac{\varepsilon-\mu}{k_B T}}\right), f(\varepsilon) = \frac{1}{1+e^{\frac{\varepsilon-\mu}{k_B T}}} \quad (2)$$

and

$$\Omega_n^z(k) = i \sum_{n'\neq n} \frac{\left\langle n\left|\frac{\partial H}{\partial k_x}\right|n'\right\rangle\left\langle n'\left|\frac{\partial H}{\partial k_y}\right|n\right\rangle - (x\leftrightarrow y)}{\left(\varepsilon_n - \varepsilon_{n'}\right)^2} \quad (3)$$

Besides, our tight binding model consists of 10 Mn 3$d$ orbitals and 6 Si 3$p$ orbitals. For doping, electrons on Mn 4$s$ orbitals are also considered since the energy is larger than that on Mn 3$d$ orbitals. Thus, the doping of $Mn_{5+x}Si_{3-x}$ induces 10$x$ extra electrons per cell, shifting Fermi levels by 0.02 eV, 0.05 eV, and 0.07 eV for $x$ = 0.04, 0.10, and 0.13, respectively.

**Phenological toy model and Monte Carlo simulation**

The anomalous Hall conductivity at zero temperature $\sigma_{xy}(\varepsilon_F)$ is assumed as the cosine function of Fermi energy $\varepsilon_F$, with an energy period $\epsilon_0$ that mainly depends on the band structure of specific material (text S2). From the equation below,

$$\alpha_{xy}(\mu, T) = -\frac{k_B}{e} \int d\varepsilon \left(-\frac{\partial f}{\partial \varepsilon}\right) \frac{\varepsilon - \mu}{k_B T} \sigma_{xy}(\varepsilon) \tag{4}$$

the temperature dependence of anomalous Nernst conductivity $\alpha_{xy}$ can be calculated as

$$\alpha_{xy} \propto \left(\frac{T}{T_P}\right) e^{-\left(\frac{T}{T_P}\right)^2} \tag{5}$$

where $T_P$ is proportional to $\epsilon_0$. Furthermore, a Mont Carlo simulation of Mn$_5$Si$_3$ was performed with the exchange energy calculated by VASP software. Inspired by the phenological result of the square lattice Ising model (*68*), the thermal fluctuation of the Néel order is fitted as,

$$N(T) = N(0) \left[1 - \sinh^{-4}\left[\ln(\sqrt{2} + 1)\frac{T_N}{T}\right]\right]^{\frac{1}{3}} \tag{6}$$

in the region $T < T_N$ = 225 K. A simple but effective way to capture the feature of temperature dependence of $\alpha_{xy}$ is to multiply Eq. 5 and Eq. 6 for considering the influence of both the band structure and the thermal fluctuation of the Néel vector.

**Acknowledgments:**

We acknowledge BL02U2 of Shanghai Synchrotron Radiation Facility and the support from Beijing Innovation Center for Future Chip (ICFC), Tsinghua University. Some devices were fabricated via an Ultraviolet Maskless Lithography machine (Model: UV Litho-ACA, TuoTuo Technology).

**Funding:**

National Key R&D Program of China (Grant No. 2022YFA1402603, 2021YFB3601301)

National Natural Science Foundation of China (Grant No. 52225106, 12241404, and 12022416)

Hong Kong Research Grants Council (Grant No. 16303821, 16306722, and 16304523)

Natural Science Foundation of Beijing, China (Grant No. JQ20010)

**Author contributions:**

L. H., W. H., Y. Z., and J. D. prepared the samples and carried out transport measurements. X. F. performed calculations. This work was conceived, led, coordinated, and guided by L. H., X. F., W. H., J. L., C. S., and F. P. All the authors contributed to the writing of the manuscript.

**Competing interests:**

Authors declare that they have no competing interests.

**Data and materials availability:**

All data needed to evaluate the conclusions in the paper are present in the paper and/or the Supplementary Materials.